\documentclass[journal=jacsat,manuscript=article]{achemso}

\usepackage{chemformula} 
\usepackage[T1]{fontenc} 
\usepackage{graphicx}
\usepackage{soul} 
\usepackage{longtable}
\usepackage{siunitx}

\author{Yusuke Tamura}
\author{Kairi Masuda}
\email{kairi.masuda.c5@tohoku.ac.jp}
\author{Yu Kumagai}
\affiliation[Tohoku University]
{Institute for Materials Research, Tohoku University, 2-1-1 Katahira, Aoba-ku, Sendai, 980-8577, Japan}

\title[An \textsf{achemso} demo]
  {{On-demand phase-field modeling: Three-dimensional Landau energy for HfO$_2$ through machine learning}}

\abbreviations{IR,NMR,UV}

\allowdisplaybreaks

\begin{document}







\begin{abstract}
The unexpected emergence of ferroelectricity in HfO$_2$ at reduced dimensions has attracted considerable attention, as it provides a pathway toward the realization of ultrasmall ferroelectric devices. Ab initio calculations suggest that this effect arises from a unique mode coupling, in which an antipolar displacement mode stabilizes a robust polar distortion. Based on these insights, Landau–Devonshire energy models have been proposed using such lattice modes as order parameters. However, most existing models are limited to a simplified one-dimensional model because of the computational cost of ab initio calculations and the limitations of conventional Landau polynomials. Here, we constructed a three-dimensional Landau–Devonshire potential for HfO$_2$ by employing the tetragonal, antipolar, and polar modes as coupled order parameters, based on the latest machine-learning technologies. We generated a large-scale dataset of energies over a three-dimensional structural space, with the computational cost drastically reduced through the use of machine-learning interatomic potentials, and trained a multilayer perceptron (MLP) to learn the relationship between the order parameters and the energy. The energy predicted by the MLP successfully captures the characteristic coupling behavior whereby the antipolar modes induce the polar mode. Furthermore, by extending this MLP-based Landau potential to a position-dependent functional, that is, to a phase-field modeling framework, we revealed that the polarization magnitude in thin films decreases compared with the bulk state, while the critical strain required for the onset of spontaneous polarization increases due to surface effects. This study presents a new framework for the on-demand construction of Landau energy and phase-field modeling using the latest machine-learning techniques, enabling multiscale analysis of complex ferroelectric phenomena.
\end{abstract}

\clearpage

\section{Introduction}

Conventional perovskite ferroelectrics such as PbTiO$_3$ have long been archetypes for switchable electronic polarization due to symmetry-lowering structural distortions\mbox{\cite{Cohen1992, isupov2002phases, shirane1952phase}}. This property has been utilized in non-volatile memories and actuators, which have been miniaturized to nanometer scales to increase the density in devices\mbox{\cite{Scott1989,Setter2006,Park2023Revival,Park2023Overcoming}}. However, as the system size approaches a few nanometers, depolarizing fields due to incomplete screening, surface recombination, and related effects suppress the long-range order, driving a size-dependent collapse of polarization and hindering their application to nanodevices\mbox{\cite{batra1972thermodynamic, mehta1973depolarization, ni2018critical, shaw2000properties}}. The recent discovery of robust ferroelectricity in nominally centrosymmetric HfO$_2$-based thin films overturns this paradigm; that is, HfO$_2$ retains polarization even at sub-10-nm scales in its non-centrosymmetric orthorhombic phase\mbox{\cite{boscke2011ferroelectricity, Muller2012Ferroelectricity, Sang2015, Polakowski2015}}. This behavior not only redefines the scaling limits of ferroelectric electronics, heralding a new generation of ultra-compact, CMOS-compatible non-volatile memories\mbox{\cite{Mikolajick2018, fan2016}}, but also extends our knowledge of the origin of ferroelectricity beyond conventional perovskites. In other words, understanding and engineering this robust ferroelectricity at reduced dimensions are scientifically and technologically pivotal for next-generation nanoferroelectric devices.

As the origin of the robust ferroelectricity of HfO$_2$, a unique coupling of displacement modes has been pointed out based on ab initio calculations\mbox{\cite{zhou2022strain,PhysRevLett.134.136802, doi:10.1073/pnas.2406316122, 10.1063/5.0180064,Zhu2024HfO2review}}. That is, by applying strain to the tetragonal phase of HfO$_2$, an antipolar mode appears, and this mode further stabilizes polar modes. In other words, the ferroelectricity of HfO$_2$ is stabilized by successive phase transitions and is thus robust even at reduced dimensions, because the antipolar mode that stabilizes the polar mode is not affected by the depolarization field due to surface charges. To further clarify and scale up these insights, Landau–Devonshire energies have been formulated for HfO$_2$, in which the energy of the system is coarse-grained into a polynomial form in terms of these displacement modes\mbox{\cite{zhou2022strain,PhysRevLett.134.136802, doi:10.1073/pnas.2406316122, 10.1063/5.0180064,Zhu2024HfO2review}}. Thus, the investigation of phase transitions in HfO$_2$ is now moving from the atomistic to the continuum-mechanics regime. However, the Landau–Devonshire energies for HfO$_2$ proposed so far are simplified one-dimensional models, that is, they only consider one-directional polarizations\mbox{\cite{zhou2022strain,PhysRevLett.134.136802, doi:10.1073/pnas.2406316122, 10.1063/5.0180064,Zhu2024HfO2review}}. This limitation restricts their applicability to phase-field modeling, that is, to the simulation of three-dimensionally inhomogeneous polarization needed to investigate the phase behavior of HfO$_2$ under more realistic conditions experienced by thin films, which typically include depolarization fields, thickness-dependent mechanical strain, and the formation of ferroelectric domains\mbox{\cite{chen2002phase,chen2008phase,ZHOU2022117920}}. Therefore, a Landau–Devonshire theory beyond one dimension is needed to further extend our knowledge of the robust ferroelectricity in HfO$_2$.

Constructing a three-dimensional Landau–Devonshire energy for HfO$_2$ is challenging for the following reasons. In previous studies, the Landau–Devonshire energy for HfO$_2$ was constructed from first-principles calculations, because the complex interplay among multiple modes in HfO$_2$ makes it extremely difficult to determine the coefficients reliably from experiments\mbox{\cite{zhou2022strain,PhysRevLett.134.136802, doi:10.1073/pnas.2406316122, 10.1063/5.0180064,Zhu2024HfO2review}}. However, extending the Landau energy to a fully three-dimensional form requires an extensive exploration of a high-dimensional order-parameter space, which in turn demands an enormous number of first-principles calculations and thus renders the computational cost practically prohibitive\mbox{\cite{Bhattacharyya2019}}. In addition, as the number of order parameters increases, the number of polynomial terms in the Landau expansion grows combinatorially, making it infeasible to construct the functional explicitly\mbox{\cite{sheng2010modified,10.1063/1.3488636,C9CP03802G}}. Moreover, there are intrinsic limitations in representing the complex mode couplings in HfO$_2$ solely by a finite-order polynomial expansion.

To overcome these limitations, we propose an on-demand phase-field concept, in which we develop a three-dimensional Landau–Devonshire energy for HfO$_2$ with tetragonal, antipolar, and polar modes as order parameters with the aid of state-of-the-art machine-learning techniques. This paper is organized as follows. First, we describe the methodology for constructing a three-dimensional Landau–Devonshire energy for HfO$_2$ using machine-learning techniques. Next, we extend the obtained Landau energy to a position-dependent functional for phase-field simulation and analyze the spontaneous polarization behavior in an HfO$_2$ thin film. Finally, we summarize the results of this study.

\clearpage
\section{Methods}
\subsection{A MLP-based Landau potential for HfO$_2$}

Figure 1 shows primary displacement modes relevant to the appearance of the ferroelectric phase. In previous studies, a successive phase-transition mechanism has been proposed for the emergence of ferroelectricity: applying uniaxial strain to the cubic phase first induces a tetragonal phase, then further strain leads to an antipolar phase, and finally this antipolar distortion induces a polar mode responsible for ferroelectricity. Based on this scenario, a one-dimensional Landau potential, that is, a Taylor expansion of the energy in terms of these displacement modes has been proposed\mbox{\cite{zhou2022strain,PhysRevLett.134.136802, doi:10.1073/pnas.2406316122, 10.1063/5.0180064,Zhu2024HfO2review}}.

To extend this Landau expression to three dimensions, it is necessary to capture the energy variations with respect to the nine modes $T_i$, $A_i$, and $P_i$ ($i = x, y, z$), as shown in Supplementary Information 1. Moreover, it is essential to incorporate the full strain tensor $\varepsilon_{ij}$, including both normal and shear components. However, as the number of order parameters increases, the polynomial form becomes increasingly complex and may fail to accurately capture the couplings among different modes. Therefore, in this study, we construct a Landau potential for HfO$_2$, $f_{\mathrm{MLP}}(T_i, A_i, P_i, \varepsilon_{ij})$, using a multilayer perceptron (MLP) as follows:
\begin{align}
f_{\mathrm{MLP}}(T_i, A_i, P_i, \varepsilon_{ij}) = W^{(L)} \boldsymbol{h}^{(L-1)} + \boldsymbol{b}^{(L)}, 
\end{align}
where $L$ denotes the total number of layers in the MLP, and $\boldsymbol{h}^{(l)}$ is the hidden representation at layer $l$ described as follows:
\begin{align}
\boldsymbol{h}^{(l)} = \sigma\!\bigl(W^{(l)} \boldsymbol{h}^{(l-1)} + \boldsymbol{b}^{(l)}\bigr),\quad l = 1, \dots, L-1, 
\end{align}
$W^{(l)}$ and $\boldsymbol{b}^{(l)}$ are the weight matrix and bias vector of layer $l$, respectively, and $\sigma(\cdot)$ is an elementwise nonlinear activation function. The input layer $\boldsymbol{h}^{(0)}$ collects all distortion modes $T_i, A_i, P_i$ and strain components $\varepsilon_{ij}$ into a single input vector, that is, $\boldsymbol{h}^{(0)} = \bigl(T_i, A_i, P_i, \varepsilon_{ij}\bigr)$. Unlike conventional polynomial-type Landau expansions, the MLP-based potential $f_{\mathrm{MLP}}$ does not assume any predefined analytical form. Instead, it learns the intrinsic nonlinear couplings among the order parameters directly from data while preserving full differentiability with respect to all inputs through automatic differentiation\mbox{\cite{baydin2018automatic}}. This differentiable nature enables seamless integration of the potential into the time-dependent Ginzburg--Landau (TDGL) equations for phase-field simulations.


\begin{figure}
  \includegraphics{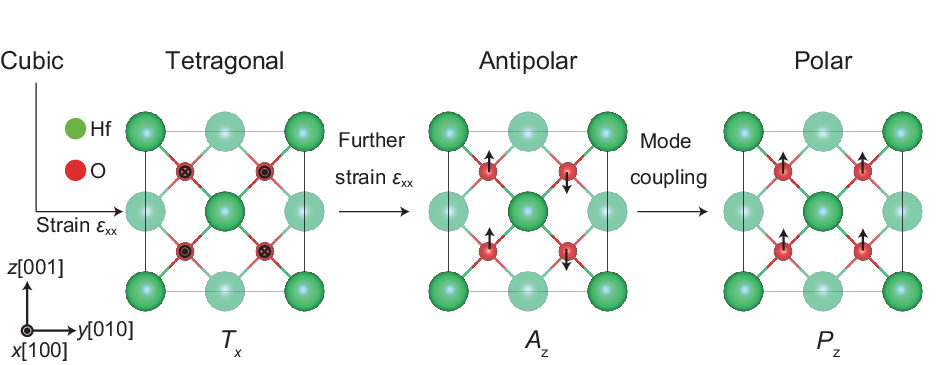}
  \caption{A schematic illustration of successive phase transitions in HfO$_2$. The arrows indicate the displacement direction in each mode.}
  \label{fig:nvttime}
\end{figure}

\clearpage

\subsection{Data generation using a machine learning potential and training the MLP}

To train the MLP, we generated a highly accurate reference dataset of energies for various HfO$_2$ configurations using a state-of-the-art machine-learning potential. Figure~2(a) schematically illustrates the data-generation procedure based on grid sampling. Because our interest lies in the phase transition from the cubic to the orthorhombic structures, we systematically explored this pathway.
The uniaxial strain component $\varepsilon_{xx}$ was set to three discrete values of 0.04, 0.05, and 0.06, while all other strain components ($\varepsilon_{yy}$, $\varepsilon_{zz}$, $\varepsilon_{xy}$, $\varepsilon_{yz}$, and $\varepsilon_{zx}$) were fixed to zero. The tetragonal mode $T_x$ was varied at 0.2, 0.3, and 0.4~\AA, whereas the other tetragonal components ($T_y$ and $T_z$) were fixed at 0 or 0.1~\AA. The antipolar modes $A_x$, $A_y$, and $A_z$ were sampled over five discrete values of 0, 0.1, 0.2, 0.3, and 0.4~\AA. For the polarization modes, $P_x$ and $P_y$ were fixed at 0 or 0.1~\AA, while $P_z$ was varied over 0, 0.1, 0.2, 0.3, and 0.4~\AA. All possible combinations of these parameters were enumerated, resulting in a total of 90,000 configurations. 
Then, we relaxed these structures while fixing the magnitudes of $T_i$, $A_i$, $P_i$, and $\varepsilon_{ij}$, and evaluated their energies, i.e., the displacement modes orthogonal to $T_i$, $A_i$, and $P_i$ were optimized by modifying forces and displacements as follows (Supplementary Information 2):
\begin{eqnarray}
F^{*}&=&(I-C^T C)F \\ 
\Delta r^{*}&=&(I-C^T C)\Delta r \\ \nonumber
\end{eqnarray}
where $I$ is the identity matrix and $C$ is a fixed mode ($T_i$, $A_i$, or $P_i$ in this case). $F$ and $\Delta r$ are the force and displacement proposed at each step of the BFGS algorithm implemented in the Atomic Simulation Environment (ASE), before applying any constraint\mbox{\cite{broyden1970convergence, fletcher1970new, goldfarb1970family, shanno1970conditioning, larsen2017atomic}}. Structures that did not satisfy the convergence criterion, that is, a force tolerance of 0.05 eV/\mbox{\AA} within 100 optimization steps, were discarded. In this way, we obtain the energy surface between the cubic and orthorhombic structures.


On the other hand, to fully explore this large structural space, it is necessary to evaluate a vast number of structures with high accuracy. However, performing conventional ab initio calculations, that is, density functional theory (DFT) calculations, is still computationally demanding for structural optimization and energy evaluation. Therefore, we utilize cutting-edge techniques that have emerged from recent materials informatics, namely universal machine-learning potentials, to relax the structures and evaluate their energies. In this work, the structural optimizations and energy evaluations are performed using MatterSim, a state-of-the-art universal machine-learning potential based on a graph neural network architecture that explicitly incorporates three-body interactions\mbox{\cite{yang2024mattersim}}.


In addition to the grid sampling, to construct a smoother and more generalizable MLP energy model, we performed additional random sampling covering a broader configuration space, as shown in Figure~2(b). In this random sampling, $\varepsilon_{xx}$ was varied within the range of 0 to 0.07, whereas all other strain components ($\varepsilon_{yy}$, $\varepsilon_{zz}$, $\varepsilon_{xy}$, $\varepsilon_{yz}$, and $\varepsilon_{zx}$) were kept at zero. All mode amplitudes, including $T_x$, $T_y$, $T_z$, $A_x$, $A_y$, $A_z$, $P_x$, $P_y$, and $P_z$, were independently and uniformly sampled using Sobol sequences within the range of 0 to 0.4~\AA, which adequately covers the displacement amplitudes encountered along the cubic to orthorhombic transition in HfO$_2$, reaching up to $\sim$0.34~\AA. This random sampling generated 150,000 structures in total, that is, we chose a number of configurations comparable to that used in the grid sampling but slightly larger because the sampling range is wider. As in the grid sampling, we optimized these structures before evaluating their energies.


The architecture of the MLP-based Landau potential, $f_{\mathrm{MLP}}(T_i, A_i, P_i, \varepsilon_{ij})$, is a fully connected feedforward neural network consisting of four hidden layers with 256 neurons each (i.e., 256×256×256×256). The Swish function was employed as the activation function for all hidden layers\mbox{\cite{ramachandran2017searching}}. On the generated dataset, the model parameters were optimized using the Adam optimizer with a learning rate of $1\times 10^{-3}$, and the mean absolute error (MAE) was used as the loss function\mbox{\cite{kingma2014adam}}. Training was performed with a batch size of 1024. The entire dataset, containing 239,821 structures, was divided into training and validation subsets in a ratio of 8:2. To ensure robust model performance and mitigate overfitting, cross-validation was performed. Training was terminated by early stopping when the validation error did not improve for 10 consecutive epochs. From the averaged convergence behavior, the optimal number of epochs was determined to be 71. Finally, the model was retrained on the entire dataset ($N = 239{,}821$) with the same hyperparameters.

\begin{figure}
  \centering
  \includegraphics[width=80mm]{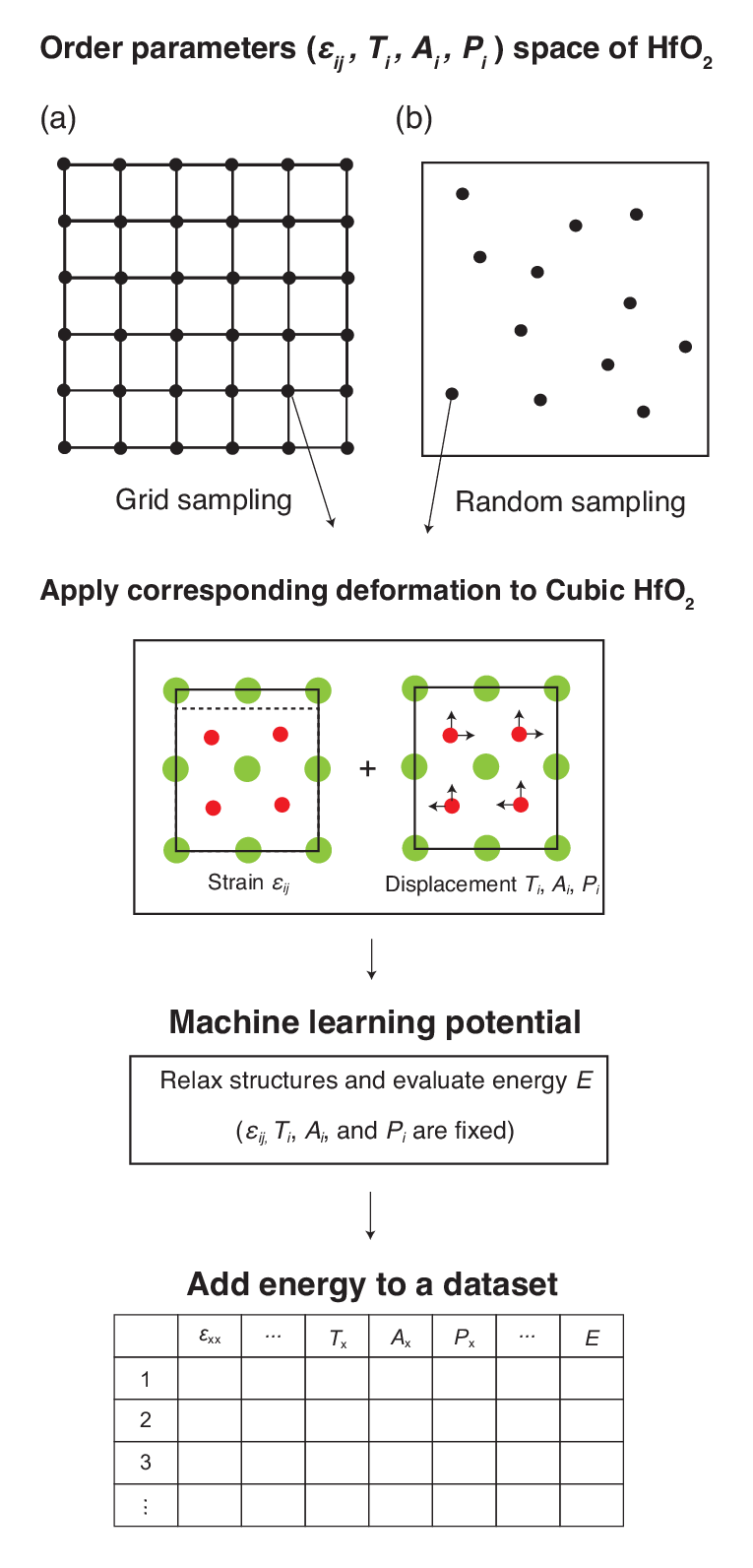}
\caption{Schematic illustration of the dataset generation procedure combining (a) grid and (b) random sampling for training the MLP.}
  \label{fig:nvttime}
\end{figure}
\clearpage

\section{Results}
\subsection{Validation of MLP-based Landau potential for HfO$_2$}

Figure 3(a) compares the energies obtained from MatterSim with those predicted by the MLP-based Landau potential.
The predicted energies are in excellent agreement with the MatterSim results, with a coefficient of determination of $R^2 = 0.9987$ and a mean absolute error (MAE) of $0.0283$ eV/atom.
These results indicate that the HfO$_2$ atomic configuration space sampled by MatterSim is successfully mapped onto the Landau–Devonshire energy landscape by our MLP-based model.
Figure 3(b) shows the potential energy surfaces associated with the tetragonal-to-orthorhombic phase transition, as obtained from both MatterSim and the MLP-based Landau potential.
As the magnitude of the antipolar displacement increases to approximately 0.27 \mbox{\AA}, an energy minimum emerges at a nonzero $P$, indicating the spontaneous stabilization of polar displacements.
This behavior demonstrates a mode-coupling phase transition, in which enhancement of the antipolar mode $A$ induces the emergence of the polar mode $P$, consistent with the strain-induced ferroelectric mechanism previously reported by Zhou \textit{et al.}\mbox{\cite{zhou2022strain}}

Next, we calculated a ground state structure by updating the order parameters in the direction that minimizes the total energy using gradient descent, with a learning rate of 0.01. The tensile strain was fixed at $\varepsilon_{xx} = 6\%$, and all order parameters were initialized to a small finite value of \mbox{\SI{1e-4}{\angstrom}}. Figure~4 shows the evolution of order parameters during energy minimization. Starting from the cubic phase, the $T_x$ mode was first activated, followed by the activation of the $A_z$ mode. Subsequently, the $A_x$, $A_y$, and $P_z$ modes were cooperatively activated. In the final equilibrium state, the $T_x$, $A_x$, $A_y$, $A_z$, and $P_z$ modes appeared, while the other modes remained nearly zero, resulting in a structure close to the ferroelectric $P2_1ca$ phase.
These results demonstrate that the developed MLP-based Landau model successfully reproduces the strain-induced stabilization of orthorhombic phase in HfO$_2$.
Therefore, we successfully extended the Landau energy model of HfO$_2$ into a three-dimensional order-parameter space.

\begin{figure}[H]
  \centering
  \includegraphics[width=90mm]{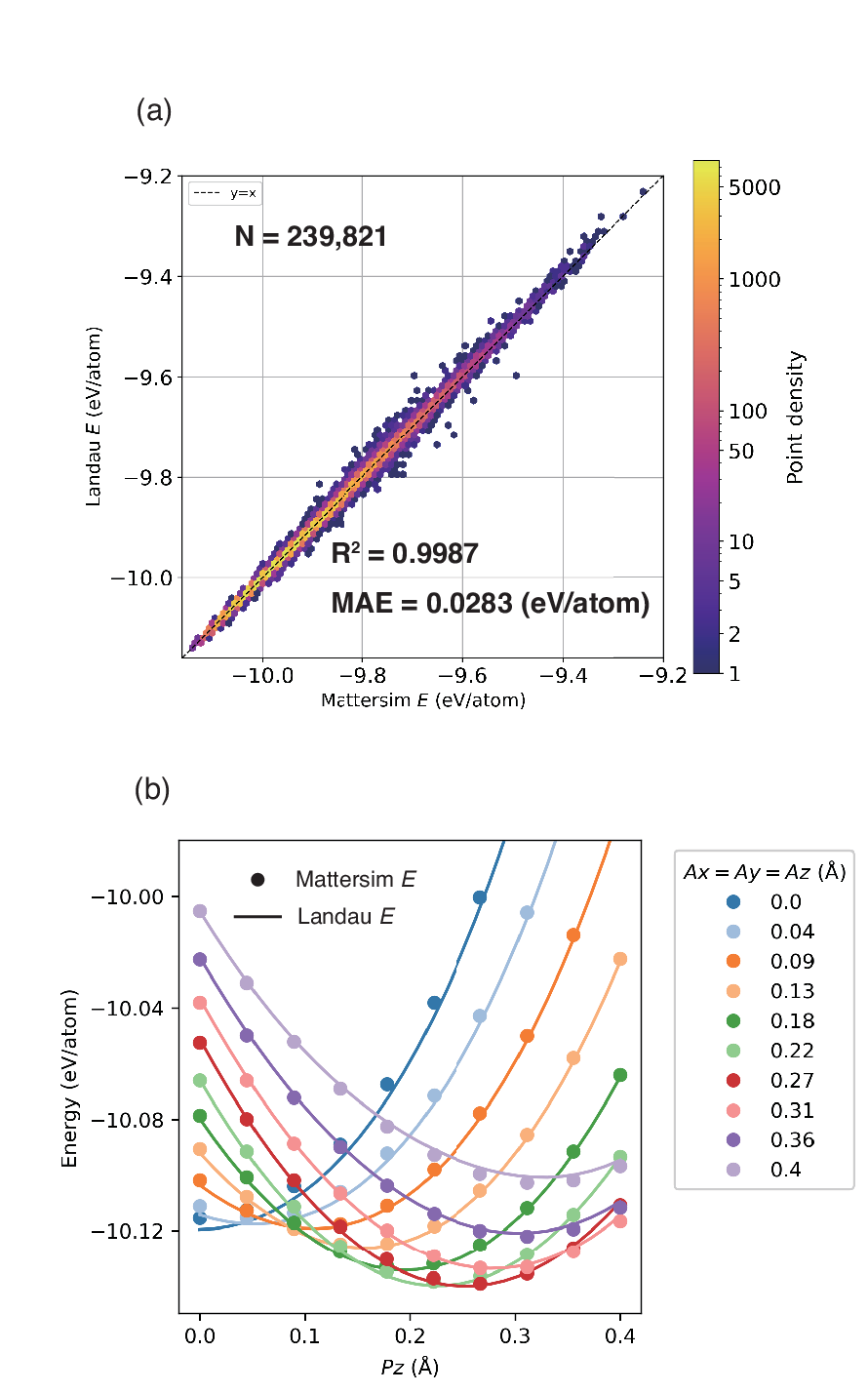}
  \caption{(a) Comparison between the energies obtained from the machine learning potential (MatterSim) and the MLP-based Landau potential using 239,821 data points employed for training. (b) Potential energy surface for the transition from the tetragonal to orthorhombic phase by varying the order parameters. The strain $\varepsilon_{xx}$ was fixed at 0.05 and the tetragonal mode amplitude $T_x$ was set to 0.3 Å, while $A_x$ = $A_y$ = $A_z$ were varied simultaneously from 0 to 0.4 Å together with $P_z$. The solid lines represent predictions from the MLP-based Landau potential, while the dots indicate the energies obtained from MatterSim.}
  \label{fig:nvttime}
\end{figure}
\clearpage
\begin{figure}
  \includegraphics[width=90mm]{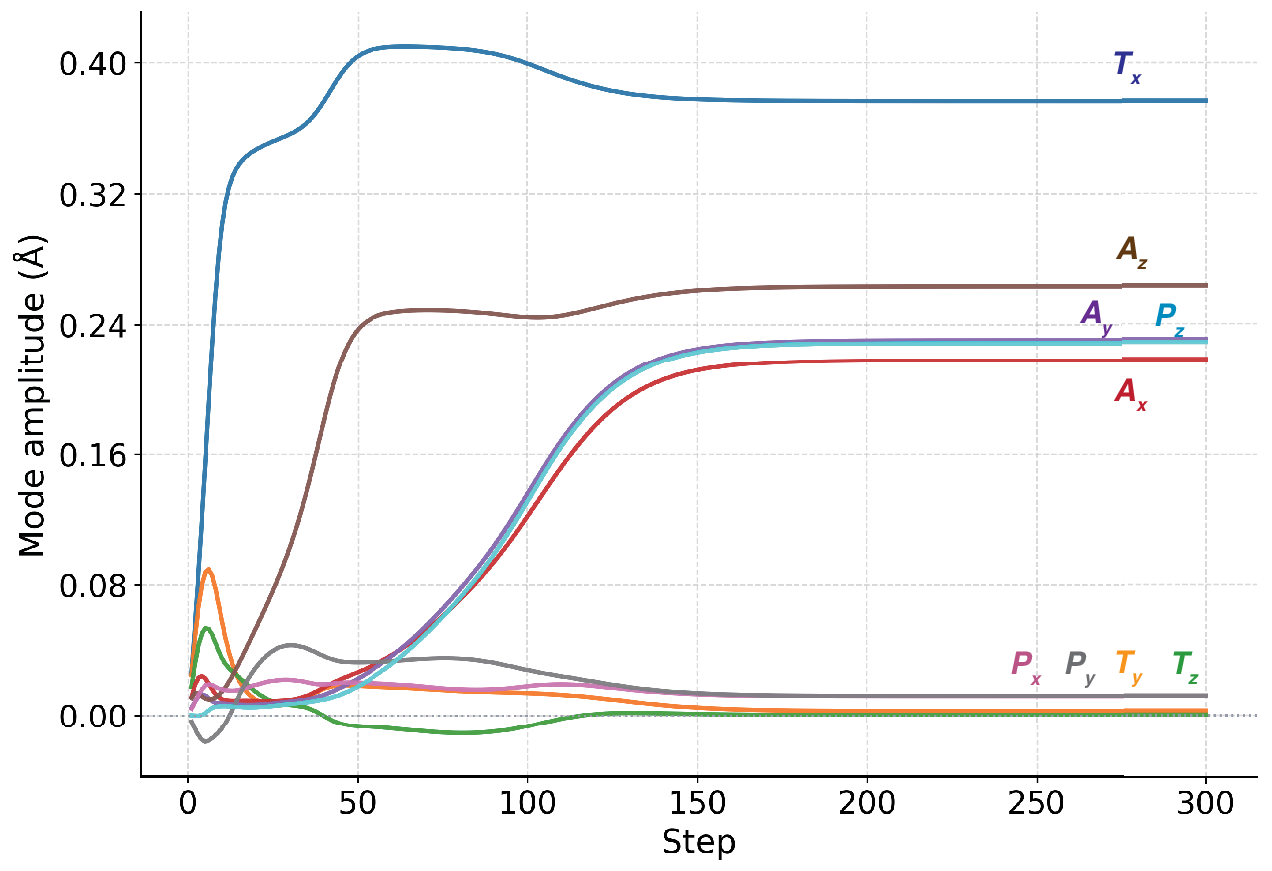}
  \caption{Evolution of the order parameters during energy minimization.}
  \label{fig:nvttime}
\end{figure}

\clearpage

\subsection{Extensions to phase-field simulation}
Since the constructed energy function accurately reproduced the ferroelectric behavior of HfO$_2$, we extended our modeling to a position-dependent functional, and investigated the polarization behavior in thin films using the phase-field method. That is, we define the total free energy as follows:
\begin{equation}
\begin{aligned}
F = \int_V f \, dv = \int_V (f_{MLP} + f_{elec}) \, dV,
\end{aligned}
\end{equation}
where $f$ and $f_{elec}$ denote the total free energy density and the electrostatic energy density. $V$ denotes the entire volume of ferroelectric material. The electrostatic energy density is given by
\begin{equation}
\begin{aligned}
f_{elec} = -\frac{1}{2} \varepsilon_0 \varepsilon_r(E_x^2 + E_y^2 + E_z^2) - (E_xP_x + E_yP_y + E_zP_z)
\end{aligned}
\end{equation}
where ${E_i}$ and $\varepsilon_{0}$ are the components of the electric field and the permittivity of vacuum, respectively, and $\varepsilon_{r}$ is the relative dielectric constant. Then, the time evolution of $T_i$, $A_i$, and $P_i$ is governed by the time-dependent Ginzburg-Landau (TDGL) equations,
\begin{equation}
\frac{\partial T_i(\mathbf{x}, t)}{\partial t} = -L_T \frac{\delta F}{\delta T_i(\mathbf{x}, t)},
\end{equation}
\begin{equation}
\frac{\partial A_i(\mathbf{x}, t)}{\partial t} = -L_A \frac{\delta F}{\delta A_i(\mathbf{x}, t)},
\end{equation}
\begin{equation}
\frac{\partial P_i(\mathbf{x}, t)}{\partial t} = -L_P \frac{\delta F}{\delta P_i(\mathbf{x}, t)},
\end{equation}
where $t$ represents time. $L_T$, $L_A$, and $L_P$ are the kinetic coefficients associated with the domain mobility of the tetragonal, antipolar, and polar mode amplitudes, respectively. $\frac{\delta F}{\delta T_i}$, $\frac{\delta F}{\delta A_i}$ and $\frac{\delta F}{\delta P_i}$ denote the thermodynamic driving forces for the evolution of tetragonal, antipolar and polar mode amplitudes.
In addition to the TDGL equations, the Maxwell equation
\begin{equation}
\frac{\partial}{\partial x_i} \left( -\frac{\partial f}{\partial E_i} \right) = 0,
\end{equation}
must be satisfied simultaneously for charge-free ferroelectric materials.

Figure~5(a) shows the bulk model, in which the ferroelectric material occupies the entire simulation cell. The simulation cell was a cube with a side length of \mbox{\SI{64}{\nano\meter}}, and the grid spacing was set to \mbox{\SI{1}{\nano\meter}}. The thin-film model shown in Figure~5(b) incorporates vacuum regions on both sides of the HfO$_2$ layer within the cell to mimic depolarization fields arising from free surfaces. In the thin-film model, a \mbox{\SI{16}{\nano\meter}}-thick HfO$_2$ layer was placed at the center, with \mbox{\SI{24}{\nano\meter}}-thick vacuum regions on both sides.
To reduce the Gibbs effect associated with the Fourier transforms of functions across the sharp interface between the ferroelectric material and the vacuum, we introduced a diffuse interface-shaped function to smear the interfacial region as follows:\mbox{\cite{Wang2013ActaMat}}
\begin{equation}
\eta(\mathbf{r}) = \frac{1}{2} \left\{ 1 - \tanh\left[ \beta \left( |\mathbf{r}| - \frac{d}{2} \right) \right] \right\},
\end{equation}
where $|\mathbf{r}|$ is the distance from the center of the thin film, $\beta$ is a positive parameter controlling the width of the surface region, and $d$ is the film thickness. In this study, $\beta$ was set to $1 \mathrm{nm}^{-1}$.

\begin{figure}
  \includegraphics[width=100mm]{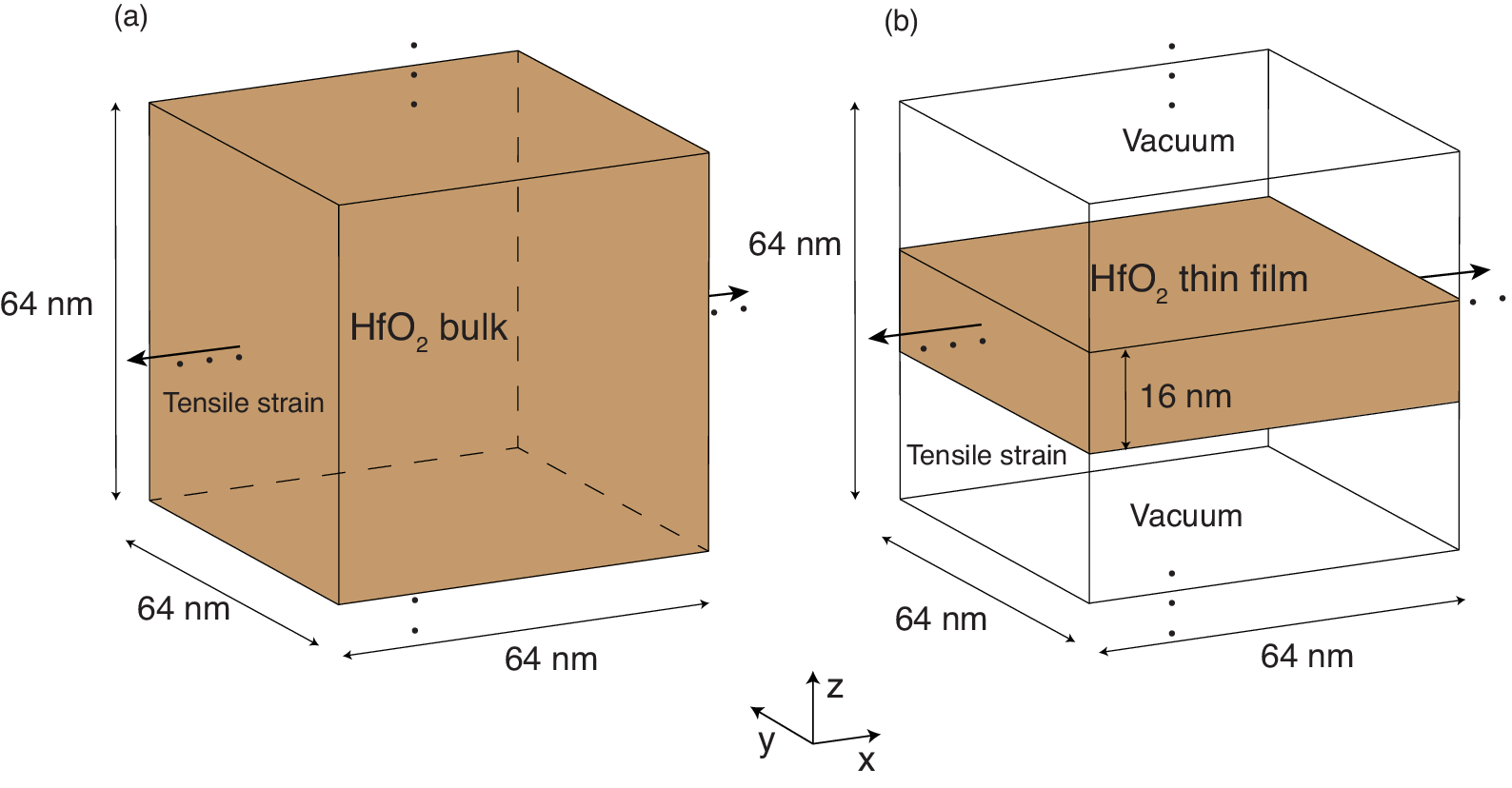}
  \caption{Schematic illustration of the simulation cells of HfO$_2$: (a) bulk and (b) thin film.}
  \label{fig:nvttime}
\end{figure}

\clearpage

To investigate the influence of surface effects on the spontaneous polarization of HfO$_2$, we simulated the thermodynamic equilibrium states of both bulk and thin-film HfO$_2$ under \mbox{\SIrange{4}{6}{\%}} tensile strain. The dielectric constant of HfO$_2$, $\varepsilon_{r}^{\mathrm{HfO_2}}$, was set to 25,\mbox{\cite{zhou2022strain}} and that of the vacuum layer, $\varepsilon_{r}^{\mathrm{vac}}$, was set to 1. The polarization magnitude, $P_{\mathrm{mag}}$ (C/m$^2$), corresponding to the $P$ modes (\AA) was estimated by a linear approximation based on the spontaneous polarization of the ferroelectric phase, that is, $P_{\mathrm{mag}} = \alpha P$, using a proportionality factor of $\alpha = \SI{1.8}{\coulomb\per\square\meter\per\angstrom}$. In the initial state, all order parameters were set to a small finite value of $T_i = A_i = P_i = \SI{1e-4}{\angstrom}$.
The time evolution of these order parameters was computed by solving the TDGL equations coupled with Maxwell’s equation, using a semi-implicit Fourier spectral method and a Fourier spectral interactive perturbation method\mbox{\cite{Chen1998CPC, Hu1998JACS,Wang2013ActaMat}}(Supplementary Information 3 and 4), with a time step of $\Delta t = \num{0.01}$. The kinetic coefficients in the TDGL equations were set to $L_T = L_A = L_P = 1$. The time integration was continued until the change in each order parameter at every time step became smaller than \mbox{\num{1e-5}} \mbox{\AA}.

\clearpage

\subsection{Strain dependence of ferroelectricity in HfO$_2$}
Figure~6(a) shows the equilibrium values of each order parameter obtained from the phase-field simulations for the bulk model, where the tensile strain was varied from 4\% to 6\% in increments of 0.1\%. In the strain range of 4.0--4.4\%, only the $T_x$ mode exhibits a large amplitude, while the $A_z$ mode is also present but remains small, corresponding to the tetragonal phase. In the subsequent strain range of 4.4--5.0\%, the $A_z$ mode starts to develop, indicating the emergence of an intermediate antipolar phase prior to the onset of spontaneous polarization. When the strain exceeds 5.1\%, the $P_z$, $A_x$, and $A_y$ modes are activated, indicating that the polar phase becomes stabilized, and the spontaneous polarization $P_z$ increases to approximately 40~$\mu$C/cm$^2$. These results demonstrate that the simulations successfully reproduce the characteristic successive phase transitions of HfO$_2$, in which the $T$, $A$, and $P$ modes emerge sequentially with increasing tensile strain.

Turning to the thin-film system shown in Figure~6(b), a similar successive activation of the order parameters is also observed. However, the emergence of the polar phase requires a larger tensile strain of 5.4\% than in the bulk case, and the magnitude of the polarization is reduced to approximately 20~$\mu$C/cm$^2$.  These results indicate that the depolarization field due to the uncompensated charge at the surfaces suppresses the development of $P_z$, reducing its magnitude and increasing the critical strain necessary for polarization onset. At the same time, these results suggest that HfO$_2$ still retains robust ferroelectricity even in the thin-film system.

In addition, focusing on the behavior of the $A$ modes in both systems, we found that while $A_z$ remains almost unchanged between the bulk and thin-film systems, the amplitudes of $A_x$ and $A_y$ are notably reduced in the thin film along with the reduction of $P_z$. This cooperative variation suggests a strong coupling between $A_x$, $A_y$, and $P_z$, which is also supported by the time-dependent evolution results shown in Figure~4. These results indicate that while the activation of the $A_z$ mode can serve as a trigger for the emergence of polarization, its behavior is independent of $P_z$. In contrast, $A_x$ and $A_y$ are stabilized by $P_z$, and in the thin-film system, the suppression of $P_z$ by the depolarization field leads to a corresponding reduction in $A_x$ and $A_y$.
The phase-transition behavior among these modes was revealed through the extension of the Landau energy model of HfO$_2$ to a three-dimensional order parameter space and the application of the phase-field method.

\begin{figure}[htbp]
  \centering
  \includegraphics[width=120mm]{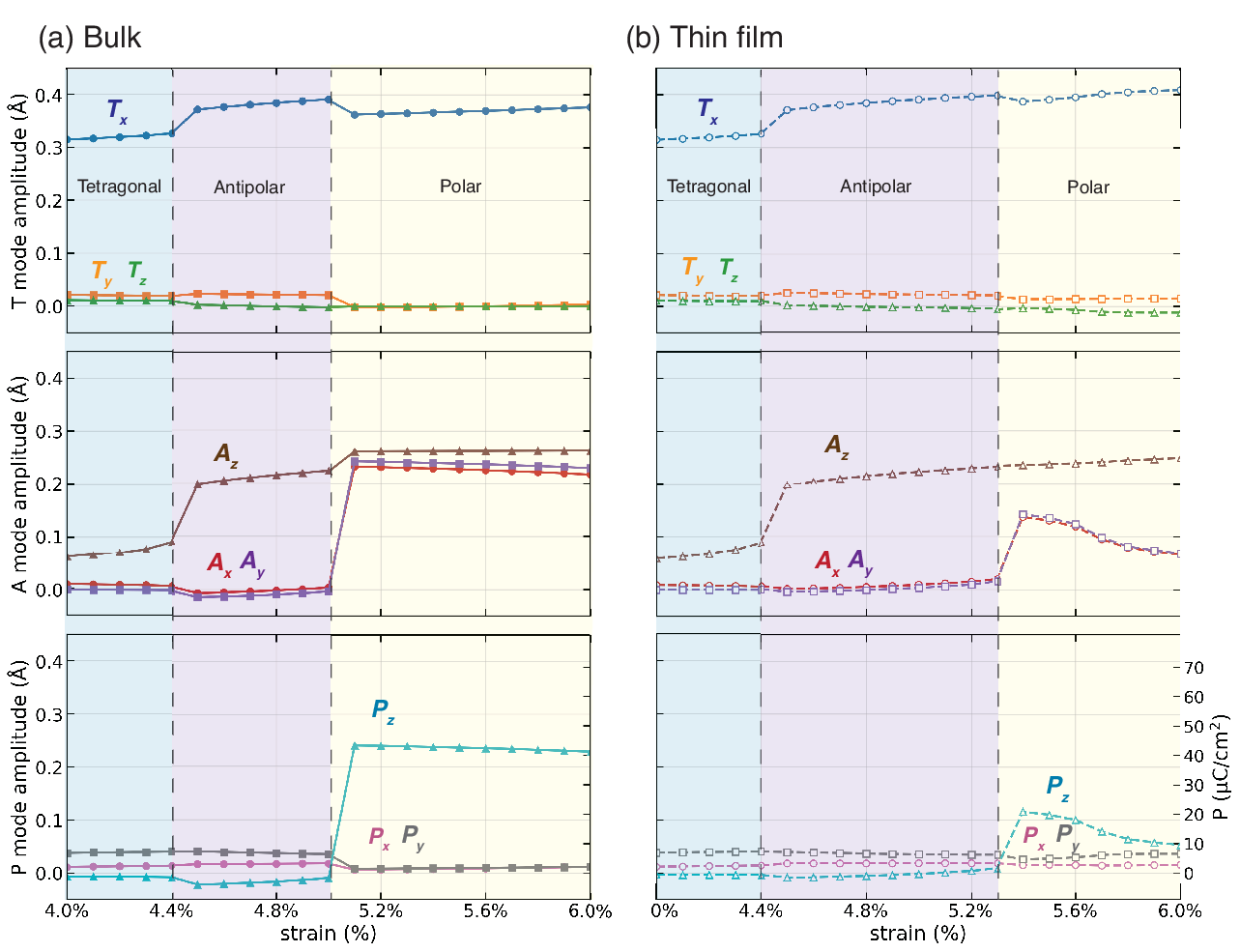}
  \caption{Representative values of the equilibrium order parameters of the $T$, $A$, and $P$ modes inside the material, obtained from phase-field simulations for (a) bulk and (b) thin-film models under tensile strains of 4--6\%. The regions colored blue, purple, and yellow represent the tetragonal, antipolar, and polar phases, respectively.}
  \label{fig:nvttime}
\end{figure}

\clearpage
\section{Discussion}
As demonstrated above, we successfully extended the Landau energy model of HfO$_2$ to a three-dimensional order parameter space. At the same time, we should mention several limitations to be addressed in future work: (1) Although the constructed energy function represents a step toward a three-dimensional extension, it currently considers only a uniaxial phase transition. However, this uniaxial model cannot treat rotational variants of the $A$ mode that appear in the presence of inhomogeneous $T$-mode distributions. Therefore, to reproduce domain-wall formation and polarization switching behavior through phase-field modeling, it is necessary to develop a symmetry-equivalent energy model that remains invariant under rotations and mirror operations, for example, by employing equivariant neural networks\mbox{\cite{Batzner2022}}. (2) In addition, to fully construct a phase-field model based on the present energy function, the gradient energy coefficients must be determined.
(3) Furthermore, the present Landau energy was formulated for a system at $0\,\mathrm{K}$, neglecting finite-temperature effects. To construct a more realistic phase-field model, it is essential to incorporate thermal contributions into the free energy.

\clearpage
\section{Conclusion}
In this study, we demonstrated an on-demand construction scheme for a three-dimensional Landau–Devonshire energy of HfO$_2$, that is, a multimode ferroelectric material, using a multilayer perceptron and machine learning potentials. We have successfully mapped the atomic configuration space of HfO$_2$ onto a Landau–Devonshire energy functional with high fidelity, and the constructed energy function accurately reproduces the strain-dependent phase transition behavior of HfO$_2$ from the cubic phase to ferroelectric phases.
Furthermore, by extending this energy functional to a position-dependent form for phase-field simulations, we revealed that, in thin films, the polarization magnitude decreases from approximately 40~$\mu$C/cm$^2$ in the bulk to 20~$\mu$C/cm$^2$, and the critical strain required for polarization increases from \mbox{\SI{5.1}{\%}} to \mbox{\SI{5.4}{\%}}. In addition, the simulations revealed that the $A_z$ mode acts as an independent trigger for the polarization $P_z$, whereas $A_x$ and $A_y$ are stabilized by $P_z$, so that depolarization-field–induced suppression of $P_z$ leads to a corresponding reduction in $A_x$ and $A_y$. 
This study provides a new framework for the on-demand construction of Landau energies using cutting-edge machine learning techniques and for their direct extension to phase-field simulations, enabling quantitative multiscale analysis of complex ferroelectric phenomena.

\bibliography{manuscript}

\end{document}